\documentclass[12pt]{article}
\usepackage{amsmath,amssymb,amsfonts}
\usepackage{adjustbox}
\usepackage{graphicx}
\usepackage{xcolor}
\usepackage{hyperref}
\usepackage{caption}
\usepackage{subcaption}
\usepackage{float}
\usepackage{mathrsfs}
\usepackage{cite}
\usepackage{url}
\usepackage{microtype}
\usepackage{booktabs}
\usepackage{authblk}
\usepackage{enumitem}

\usepackage[margin=1in]{geometry}

\title{Symmetries, Scaling Laws and Phase Transitions in Consumer Advertising Response}

\author{Javier Marín}
\affil{javier@jmarin.info}

\date{March 5, 2024}

\begin{document}

\maketitle
\begin{abstract}
Understanding how consumers respond to business advertising efforts is essential for optimizing marketing investment. This research introduces a new  modeling approach based on the concepts of symmetries and scaling laws in physics to describe consumer response to advertising dynamics. Drawing from mathematical frameworks used in physics and social sciences, we propose a model that accounts for a key aspect: the saturation effect. The model is validated against commonly used models, including the Michaelis-Menten and Hill equations, showing its ability to better capture nonlinearities in advertising effects. We introduce new key parameters — Marketing Sensitivity, Response Sensitivity, and Behavioral Sensitivity — that offer additional insights into the drivers of audience engagement and advertising performance. Our model provides a rigorous yet practical tool for understanding audience behavior, contributing to the improvement of budget allocation strategies.

\vspace{0.5em}
\noindent\textbf{Keywords}: consumer response curve, advertising measurement, universality principle, phase transitions, statistical physics, marketing dynamics, media mix modeling, 
\end{abstract}

\section{Introduction}

The dynamics of consumer response to advertising is complex and has been studied extensively. To better understand these processes, we draw on theoretical frameworks from disciplines such as physics and social psychology. In this context, fundamental principles from physics—such as symmetries, scaling laws, and phase transitions—offer a promising approach to modeling consumer behavior. The advantage of applying the mathematical formalisms of these fundamental laws lies in their simplicity. When physicists have tried to describe a natural phenomenon, they have done so by excluding everything that could be considered accessory until they found the most basic mechanisms. Thus, they have discovered that at this very basic level, mathematical formalisms used to describe these laws, in addition to their simplicity, are very similar to each other. This is what makes these formalisms and their applications in other fields attractive. Furthermore, insights from social psychology, particularly in areas like attitude change and social influence, provide a formal basis for testing these mathematical analogies applied to consumer response dynamics.

\setlength{\parskip}{12pt}
Symmetries are one of the most fundamental principles of the physical reality that surrounds us \cite{Conway2016, Glattfelder2019}. They provide valuable frameworks for understanding equilibrium states and interactions in complex systems. While consumer behavior differs from physical systems, there may be value in adapting symmetry formalisms to analyze advertising response mechanisms.
Consumer response curves show patterns that suggest potential underlying symmetries and—perhaps more importantly—critical points where these symmetries break. The traditional S-curve model shows approximate point symmetry around its inflection point, which could represent a conserved property within specific regions of the response curve. However, real-world consumer behavior introduces asymmetries at critical thresholds that may be mathematically analogous to phase transitions in physical systems.
The saturation effect in advertising, where spending reaches a critical point beyond which returns diminish, creates an asymptotic boundary similar to a phase transition that breaks symmetry. Similarly, threshold effects at lower advertising levels introduce asymmetries that may represent a type of phase transition. 

\setlength{\parskip}{12pt}
Current quantitative marketing models often rely on stochastic approaches that assume ergodicity \cite{Palmer1982}, which may not accurately capture the path-dependent nature of consumer response. By identifying and analyzing symmetry-breaking points in consumer response curves, we may develop more descriptive models that better model consumer behavior. This adaptation of physical formalism to marketing contexts—not as direct application but as inspired mathematical formalisms—represents an interesting path for research to study consumer behavior.

\section{Current models for consumer response}

Although widely explored in the literature, there is a lack of theoretical and empirical work providing a comprehensive model of consumer behavior in response to advertising. The initial modeling methods relied on probabilistic approaches \cite{Hauser1982} and classical econometric approaches rooted in utility theory, such as the Marshal's model \cite{Friedman1949, Prest1949, Working1927}. Marshall suggested that the equilibrium for consuming a commodity $x$ occurs when the marginal utility of $x$, $ux$ equals the product of the price of $x$, $px$ and the marginal utility of money $um$ \cite{Alford1956}:

\begin{equation}
ux = px \times um
\end{equation}

where the marginal utility of money $um$ is assumed to be the inverse of money income. As money income follows a scaling law relationship, the dependence of the marginal utility $ux$ with price $px$ also follows a scaling law.

In the curves shown in Figure 1 the individual does not consider $x$ to be important, and his utility function is additive, resulting in the given marginal utility curve $U_1x$ for $x$. Two points on the demand curve are determined by multiplying two prices, $p_1x$ and $p_2x$, by the constant marginal utility of money $u_1m$. At the positions where $u_1x = p_1x$ and $u_1m$ and $u_1x = p_2x$, the values $q_1x$ and $q_2x$ are determined, which represent two points on the demand curve $D_1x$.

\begin{figure}[h]
\centering
\includegraphics[width=0.55\textwidth]{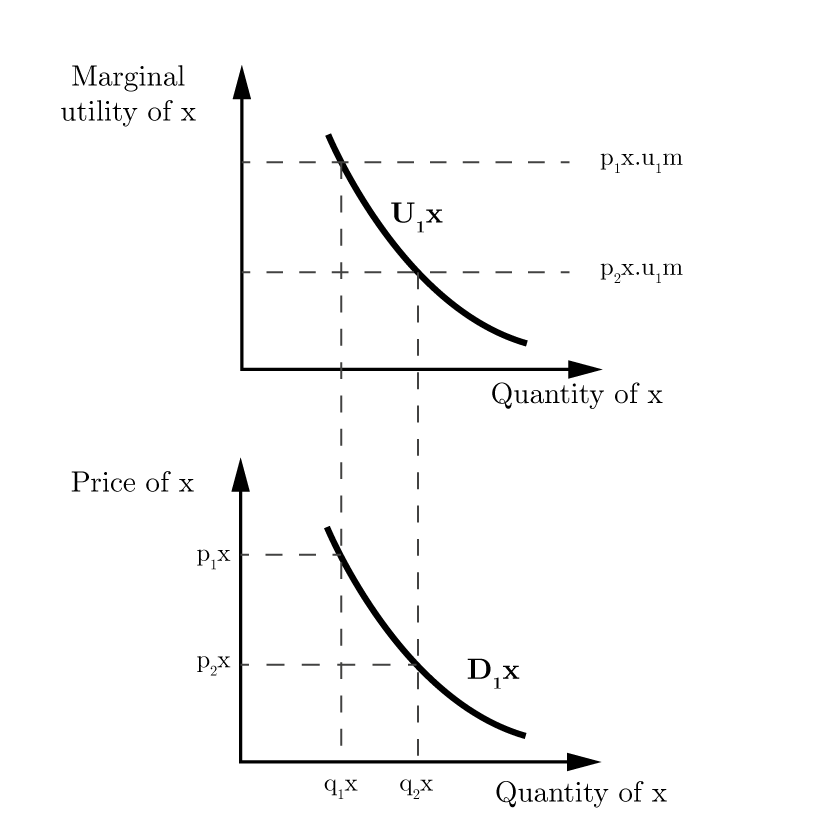}
\caption{Utility and demand curve. Adapted from Alford, 1956.}
\label{fig:utility-demand}
\end{figure}

This curve has implicit the concept of elasticity that implies a reduction of demanded quantity $qx$ when the price $px$ increases \cite{Alford1956}. The demand curve $Dx$ will follow a scaling law expression with a negative scaling parameter:

\begin{equation}
qx = Cpx^{-\alpha}
\end{equation}

Early research studies were founded on the theory of consumer information, suggesting that consumers may assess the utility of a product after making sufficient purchases of it. Advertising can inform consumers about a product, prompting them to respond and anticipate gaining an advantage they wouldn't have otherwise. Advertising helps consumers by guiding them towards optimal purchases based on their preferences, and the demand elasticity for a particular product rises post-advertisement compared to pre-advertisement \cite{Nelson1975}. This model can be represented as:
\begin{equation}
\frac{\partial Q}{\partial A} = \epsilon \frac{\partial C_A}{\partial A}
\end{equation}

where $\frac{\partial Q}{\partial A}$ is the partial derivative of quantity demanded with respect to a change in advertising, $\epsilon$ is the ordinary elasticity of demand, and $\frac{\partial C_A}{\partial A}$ is the partial derivative of costs of advertising with respect to a change in advertising. This model is a descent from econometric models that evaluate product demand based on its utility. The consumer response curve to advertising is also referred to as shape effect, particularly within the context of Marketing Mix Modelling \cite{Tellis2006}. Shape effect refers to the variation in sales as advertising intensity increases within the same time frame, and has implicit the concept of elasticity. For example, Tellis \cite{Tellis2006} defined advertising elasticity, or the elasticity of sales to advertising, as the percentage change in sales resulting from a 1\% change in advertising.

\begin{figure}[h]
\centering
\includegraphics[width=0.45\textwidth]{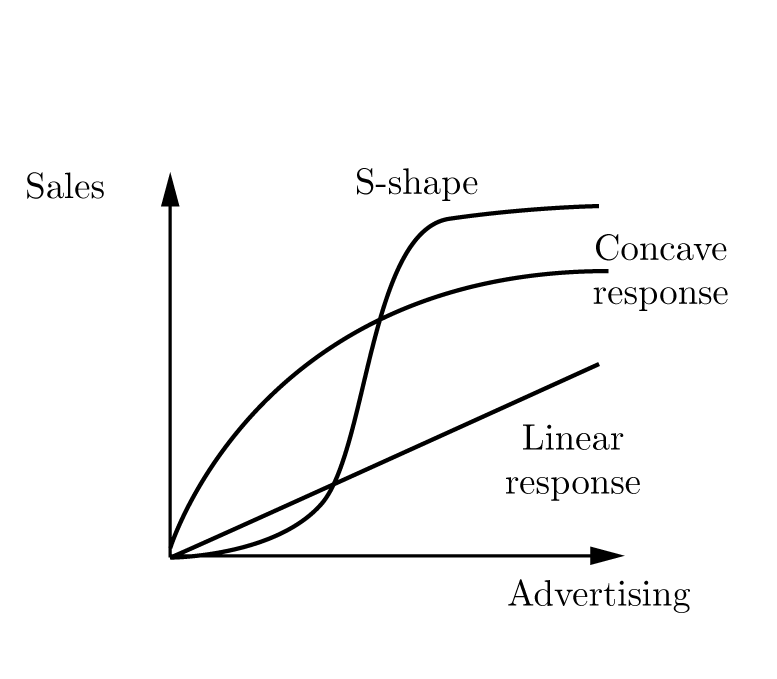}
\caption{Linear and nonlinear response to advertising}
\label{fig:responses}
\end{figure}

Shape effect can show different response curves or shapes, as illustrated in Figure \ref{fig:responses}. The representation shows three different dynamics: a linear dynamic with a fixed slope, a concave response following a scaling-law form, and an S-shaped curve. At low advertising levels in the S-shaped curve, response is modest because it is overpowered by market noise. High volumes of advertising lead to lower reaction owing to saturation. Sales show the highest responsiveness at moderate advertising levels. The S-shaped dynamic has the highest level of acceptance among the three, with the concave response curve coming in a close second. Both models share the characteristic that consumer response reaches "saturation", where it stops increasing at a given spending level. This is a central subject for all consumer response to advertising models. The S-shape curve also implies the existence of an advertising threshold at which advertising fails to create any sales response. Researchers studying shape effect analysis have explored the advertising threshold as a potential rationale for the S-shaped curve observed in Figure \ref{fig:responses} \cite{Bemmaor1984, Hanssens2003, Huang2012, Johansson1979, Simon1980, Tull1986, Vakratsas2004}. Empirical models of advertisement dynamics include thresholds that lead to concave response curves. This partially corroborates an S-shaped curve as proposed by Dubé et al. \cite{Dube2005} and \cite{Feinberg2001}. However,researchers have found the existence of a threshold in customer response to be highly improbable in reality. Vidale and Wolf \cite{Vidale1957} proposed a model that included some important assumptions like that sales rate increases with advertising rate, the existence of a saturation effect, and also that sales fluctuate with time \cite{Vidale1957}. They represented their empirical approach with the following equation:

\begin{equation}
\frac{dy}{dt} = \rho x\left(1-\frac{y}{m}\right) - \lambda y
\end{equation}

where $y$ is the sales rate, $x$ is the advertising rate, $\rho$ is the response constant, $\lambda$ is the decay constant, and $m$ is the saturation sales rate. This first-order differential equation, for a steady-state response to a constant advertising rate, has the following solution:

\begin{equation}
y = \frac{m\left(\frac{\rho x}{\lambda m}\right)}{1+\frac{\rho x}{\lambda m}}
\end{equation}

For an impulse response, the solution for the general Vidale and Wolf equation solution can be written in the following form:

\begin{equation}
\Delta y(t) = y(t) - y(0)e^{-\lambda t} = [m - s(0)][1 - e^{-\rho x/m}]e^{-\lambda t}
\end{equation}

The steady state solution exhibits a concave shape, while the spike input solution displays a convex shape. Both strategies consider a saturation effect, however neither of these solutions considers a threshold in the response. Although it was considered a highly comprehensive empirical model, it faced criticism for not accounting for the S-shaped dynamics in the response \cite{Little1979}. The discussion of whether the response curve must adhere to an S-shape or another form of shape originated with the introduction of the initial empirical models \cite{Bass1969, Johansson1973}. The two currents in the discussion agreed on the necessity of the saturation effect but disagreed on whether the model should include a possible threshold effect.

The S-curve model was originated by analyzing the derivative of the response function (sales or market share) in relation to advertising \cite{Johansson1979}. According to this approach, as advertising effort increase, the slope of the curve transitions from positive and increasing to positive but decreasing. By using this approach, the response curve can be modeled as a logistic function, where the inflection point consistently falls at the midpoint between the starting point and the saturation level. Other authors have been proposing simpler models that encompass both, the saturation effect and the threshold. One of the first proposals was made by Little \& Lodish \cite{Little1969} with a simpler equation with the advantage that does not include a specific saturation parameter but captures both effects:

\begin{equation}
r(x) = r_0 a(1 + e^{-bx})
\end{equation}

where, $x$ is the exposure level, $r$ is the return, $r_0$ is the return without advertising $r|x = 0$, and $a, b$ are non-negative constants. This approach can be understood as a conditional expectation of the average fraction potential realized for a set of consumers at exposure level $x$, denoted by $r(x)$. More recent research is using the Hill equation to model this S-shape curve \cite{Jin2017}. This equation, that has its origins in physiology \cite{Hill1910} has the flexibility of capturing both, the concave and the sigmoid shape, in the response curve.

In summary, the research on consumer response models in marketing has reached a unanimous view that customer response is saturated. This concept is pervasive throughout many fields of study, yet it should be acknowledged that there is limited empirical evidence to substantiate it. Nevertheless, to this day, there remains a lack of unanimous agreement regarding the presence or absence of a threshold in the responses. Consequently, there is ongoing discourse regarding whether the curve exhibits a concave or sigmoidal shape.
\section{Symmetries, phase transitions and the universality principle}

The origins of the universality principle can be traced back to the study of universal objects and phenomena that show common features across different mathematical contexts \cite{Batterman2017, Kadanoff1971, Stanley1971}. One of the earliest manifestations of universality can be found in number theory, where prime numbers demonstrate remarkable regularities despite their seemingly random distribution. The Riemann zeta function and its connection to the distribution of primes exemplify this universality, as evidenced by the Riemann hypothesis and its far-reaching implications \cite{Katz1974}. In probability theory, universality manifests in the behavior of random matrices, which display similar statistical properties regardless of their specific entries. This phenomenon, known as universality in random matrix theory, has been extensively studied and has profound connections to various areas of mathematics and physics, including quantum mechanics and statistical physics.

Mathematically, the universality principle can be formulated in terms of limit theorems, symmetries, and invariant structures. Limit theorems, such as the central limit theorem in probability theory, illustrate how certain statistical properties emerge universally as the size of a system grows large. Mathematically, the central limit theorem (CLT) can be expressed as follows: let $X_1, X_2, \ldots, X_n$ be independent and identically distributed random variables with mean $\mu$ and standard deviation $\sigma$. Define the sample mean $\bar{X}_n = \frac{1}{n} \sum_{i=1}^n X_i$. Then, as $n$ approaches infinity, we have the following equality:

\begin{equation}
\lim_{n\to\infty} P\left(\frac{\bar{X}_n - \mu}{\sigma/\sqrt{n}} \leq x\right) = \Phi(x)
\end{equation}

where $\Phi(x)$ is the cumulative distribution function of the standard normal distribution. This formula indicates that the standardized sample $\frac{\bar{X}_n-\mu}{\sigma/\sqrt{n}}$ converges in distribution to a standard normal random variable as $n$ becomes large. Regardless of the specific distribution of the individual random variables, the CLT ensures that the distribution of their sum or average approaches normality under certain conditions, facilitating probabilistic analysis and inference in various fields.

Symmetry principles in physics reveal underlying mechanisms that govern diverse physical phenomena, leading to universal laws and conservation principles \cite{McWeeny2002}. An important concept associated with symmetry is invariance, which refers to the fundamental property of a system to remain unchanged when subjected to manipulations. The mathematical structure that underlies the study of symmetry and invariance is known as group theory. In physics, group theory is a powerful instrument that uses symmetry to understand and predict physical phenomena \cite{Hamermesh2012, Slansky1981, Tinkham2003}.

Mathematically, transformation groups- group of transformations that do not modify a particular object or system- allow us to study symmetry. These transformations form a group by satisfying the axioms of closure, associativity, identity and inverses, representing symmetries inside the system. A group $G$ is described as a set plus an operation that joins any two elements of the group. A group action on a set $X$ is formally defined as a function $\Phi$:

\begin{equation}
\Phi: G \times X \to X
\end{equation}

where $G$ is a symmetry group if its group action $\Phi$ preserves the structure on $X$. We can explore the symmetry principle by examining rotational symmetry in classical mechanics \cite{Schattschneider1978}. Let's consider a physical system, like a rigid body, that has rotational symmetry. The rotational symmetry group for this system includes all rotations that do not alter the system's properties. This group is represented mathematically as $SO(3)$, which stands for the special orthogonal group in three dimensions. The elements of the $SO(3)$ group are rotation matrices that express rotations about an axis passing through the origin. These matrices meet the following requirements: the composition of two rotations is also a rotation (closure), rotations can be composed in any order (associativity), the identity element corresponds to no rotation (identity), and where each rotation has an inverse rotation that cancels its effect (inverses). The symmetry principle implies that if $\mathcal{L}$ represents the laws governing the physical system, then under the action of rotation $R$, the laws should remain unchanged. In mathematical terms, this can be expressed as:

\begin{equation}
\mathcal{L}(Rx) = \mathcal{L}(x)
\end{equation}

where x represents the state of the system and $Rx$ denotes the transformed state of the system under the rotation $R$.
Topological invariants are quantities associated with a topological space that remain unchanged under certain continuous transformations, such as deformations, stretching, or bending \cite{Carlsson2009}. These invariants capture essential geometric properties of the space that are independent of specific geometric configurations or coordinate systems. One of the most fundamental topological invariants is the Euler characteristic $\chi$ of a topological space $M$ \cite{Bittner2004}. It is defined as the alternating sum of the dimensions of its homology groups $H_i(M)$:

\begin{equation}
\chi(M) = \sum_{i=0}^{\infty} (-1)^i dim(H_i(M))
\end{equation}

The dimension of $H_i(M)$ represents the number of independent $i$-dimensional holes in the space, and the alternating signs ensure that the contributions of even and odd-dimensional homology groups are appropriately weighted in the sum while captures the cancellation of certain topological features, leading to a net measure of the space's Euler characteristic. These invariants remain unchanged under certain continuous transformations of the space, providing a universal perspective on the geometric properties of a space. These transformations include deformations, homeomorphisms, and continuous mappings that preserve topological properties such as connectivity, compactness, and orientability \cite{Carlsson2009}. Consider the sphere $S^2$ and the torus $T^2$ as examples. Both have different shapes and topological properties, yet they share the same Euler characteristic ($\chi = 2$). This proves how topological invariants provide a universal characterization of geometric properties that is independent of specific shapes or configurations.

\subsection{Symmetries and invariant structures}

Universality suggests the existence of underlying symmetries and principles governing phase transitions, irrespective of the specific nature of the system. Symmetry-breaking phenomena, such as spontaneous magnetization in ferromagnetic materials, play a crucial role in driving phase transitions and determining critical behavior \cite{Castellani2003}. These transitions are abrupt changes in the state of a physical system, such as the transition from liquid to gas or from ferromagnetic to paramagnetic states in magnets. Critical phenomena occur near these phase transitions, characterized by divergent behavior of certain physical quantities, such as correlation length and susceptibility. The correlation length is the scale at which the general properties of a material begin to deviate from its bulk properties. It is the distance over which there is a significant correlation between the fluctuations of the microscopic degrees of freedom (such as the positions of atoms).

Universality categorizes different physical systems based on their behavior near critical points, regardless of their specific microscopic details. This leads to the emergence of universal scaling laws (also called power laws), which describe how physical quantities behave as the complex system approaches criticality. Scaling laws can be seen as laws of nature describing complex systems. Complex systems are made up of interconnected elements that interact to create an emergent structure. The scaling hypothesis asserts that near critical points, physical quantities in complex systems exhibit a scaling behavior can be described using power laws \cite{Amaral1998, Plerou2004, Spurrett1999, West2005}. For example, the correlation length $\xi$ near a critical point $T_c$ can be modeled as:

\begin{equation}
\xi \sim |T - T_c|^{-\nu}
\end{equation}

where $\nu$ is the critical exponent characterizing the divergence of the correlation length. This expression suggests that the characteristic length scale $\xi$ diverges as the $T$ approaches the critical point $T_c$ following a power-law behavior with a negative exponent $-\nu$. During phase transitions, such as the change from a liquid to a gas or the shift between ferromagnetic and paramagnetic states, particles in the system become interrelated. The correlation length quantifies the extent to which correlations between particles endure across a certain distance. The system's correlation length tends to grow significantly when it gets closer to critical events around a phase transition point \cite{Gorban2021, Laidler1983}. As the system gets closer to the critical point, the correlation length increases, a phenomenon known as a divergence. A characteristic of critical phenomena and phase transitions is the divergence of the correlation length, which indicates the formation of long-range correlations and collective behavior in the system.

A scaling law can be described as a basic polynomial functional relationship of the type
\begin{equation}
y = Cx^{\alpha}
\end{equation}
parametrized by a scaling parameter $\alpha$, which can be positive or negative, and a constant $C$. A change in the quantity $x$ leads to a corresponding change in the quantity $y$, regardless of their initial value. In other words, changing the input value $x$ does not alter the shape of the function's output $y$. This property is known as scale invariance since it holds true for all scales \cite{Glattfelder2019}. Its mathematical expression is the following:
\begin{equation}
y = f(ax) = C(ax)^{\alpha} = a^{\alpha}f(x) \sim x
\end{equation}

Scaling-law distributions shows very large occurrences, and observations span a wide range of magnitudes. The probability density function of these distributions can be expressed as
\begin{equation}
p(x) = Cx^{-\alpha} \text{ for } \alpha > 0
\end{equation}

If $p(x)$ follows a scaling law distribution, the cumulative distribution function $P(x)$ also follows a power law, with an exponent $\alpha - 1$. Scaling law distribution can be observed in a wide range of real-world complex systems like, for example, the size and population of cities or the price moves in financial markets. Scaling laws can also apply across various physical systems undergoing phase transitions, including magnets, fluids, and cosmological models. Despite the differences in the microscopic details of these systems, they exhibit similar scaling behaviors near critical points.

\subsection{Partition function and phase transitions}

Partition functions are basic quantities in statistical physics that are used to characterize a system's thermodynamic properties \cite{Mandl1991}. This function provides details about the organization of energy levels and the probability of various particle arrangements. Partition functions frequently incorporate exponential terms that represent the Boltzmann factor, which quantifies the likelihood of occupying a specific energy state. Consider a system with a fixed volume $V$ and composition $N$. We assume that the system is in thermal equilibrium with a large reservoir at a specific temperature $T$. The weighted partition function of a system \cite{Moyal1949} can be formalized as follows:

\begin{equation}
Z = \sum_i e^{-\frac{E_i}{kT}}
\end{equation}

where $i$ is the system micro-state, $E_i$ is the energy of the state $i$, $T$ is the temperature, and k is the Boltzmann factor. The term $e^{-\frac{E_i}{kT}}$ is frequently known as the Boltzmann factor and can be considered as a weighting factor for each independent micro-state of assembly. We can rewrite the Boltzmann factor as $e^{-\beta E_i}$. This factor establishes a connection between the microscopic domain of mechanics and the macroscopic world of thermodynamics by relating the energy of the molecules in a system to the temperature of the surrounding environment \cite{Battaglia2009}. Boltzmann factor essentially gives us the probability of finding the system in a particular configuration based on its energy and temperature.

\section{Social dynamics of consumer response to advertising}

Statistical physics has been already applied to social dynamics by some researchers \cite{Helbing2010, Jusup2022, Weidlich1991}. Castellano et al., \cite{Castellano2009} extensively examined opinion dynamics, voter models, majority rule models (MR), and social impact theory using the same principles described in section 2. Social dynamics replaces the concept of physical particles with individuals. Individuals interact with each other in a complex behaviour, from which emerge patterns, sometimes not easily visible, that describe the behavior of the entire system (whether it is a small group or a broader social context). However, we haven't found references in bibliography applying these basic principles in analyzing consumer response to marketing stimuli. The field of marketing can be reviewed through the lens of social influence theory, although it has a particularity: individuals interact with a specific advertising campaign while also interacting with other individuals when developing an attitude towards a brand. Actually, this dynamics closely resembles the Ising model \cite{Glauber1963}, describing how a ferromagnetic particle tends to align its spin value with that of its neighboring particles, while also allowing for fluctuations based on temperature changes. 

However, there are other decisive factors in the relationship between consumers and advertising. The preference of a product by a social group improves the transition from exposure to engagement. The intensity of an individual's affiliation with a social group influences this shift even more, especially when the brand is well aligned with the group's beliefs \cite{Farivar2022, Spears2011}. As individuals seek approval within their social circles, social validation works as a catalyst, driving the transition from passive exposure to active participation \cite{Brown2019}. Group identity dynamics \cite{Chen2009} influence how individuals allocate attention as they move from attracting attention to real conversion. Within-group social comparison, together with in-group affinity \cite{Wachter2020}, play a key role in evaluating the possibility of attention turning to conversion. In the transition from generating interest to making a purchasing choice, social proof becomes a guiding force as people look to the activities of others in their social circle \cite{Karasawa1991}. The level of group cohesion, which is supported by shared preferences \cite{Greer2012}, influences the transition rate, with strong group relationships supporting collective decision-making. Finally, social identity theory \cite{Charness2020, Foroudi2019, Karasawa1991} emphasize the need of matching a product to an individual's social identity in order to accelerate the transition from discovery to purchase.

When proposing a new model to describe the underlying dynamics of consumer response, we will rely on the fundamental principles described in the previous section, considering as well some specific social dynamics. As we have already mentioned, a central concept for consumer response to advertising found in the literature is that consumer response tends to saturate. In our proposal, we need to consider this effect. Regarding the idea of advertising threshold, as we have seen in empirical research, there is a significant challenge in practically verifying its existence. On the other hand, the shape of the consumer's response curve will have to be drawn by the presence of regularities and either local or global symmetries in the system. The well-known S-shape is initially ruled out for two fundamental reasons: first, the presence of an advertising threshold is very difficult to prove, and second, due to the inherent limitations of this curve in real dynamics. Although these curves have been widely used in econometrics and ecology (for example, Malthusian growth), their theoretical development has proven to be very limited when analyzing complex systems.
\section{A new equation for consumer response curve}

Based on these premises, we propose a model that can be represented by the following equation:
\begin{equation}
y = Cx^{\alpha}(1 - e^{\beta x})^{-\gamma}
\end{equation}

where $y$ represents the consumer response, $x$ represents the advertising spend, $C$ is a constant reflecting the overall effectiveness of the advertising campaign, $\alpha$ is the scaling parameter characterizing the relationship between consumer response and advertising spend, $\beta$ represents the rate of change in consumer response with respect to advertising spend that incorporates the concept of symmetry-breaking phenomena, and $\gamma$ is the critical exponent that captures the behavior near a phase transition in consumer response. Next, we will provide a rationale for each of these parameters:

\begin{itemize}
\item The term $Cx^{\alpha}$ is the same than in Equation 13, introducing the property of scale invariance, meaning that varying the value of the consumer response function argument $x$, the shape of the function $y$ is preserved. Alpha, $\alpha$ represents the scaling parameter that influences the overall shape and magnitude of the consumer response curve. It implies that a change in the quantity $x$ leads to a corresponding change in the quantity $y$, regardless of their initial sizes or changing the input value $x$ does not alter the shape of the function's output $y$. Although this parameter does not directly illustrates how advertising investment affects sales, it does determine how steep or flat the curve is. A higher $\alpha$ value indicates a steeper curve, where small changes in advertising spend lead to significant changes in consumer response, while a lower $\alpha$ value corresponds to a flatter curve, where larger changes in advertising spend are needed to produce noticeable effects on consumer behavior. It is important to clarify that this parameter is different from the channel influence obtained by current MMM models. The parameter $\alpha$ can take on values ranging from 0 to 1. Advertising response $y$ tends to grow linearly with spend $x$ when $\alpha = 1$, and it is constant and equal to $C$ when $\alpha = 0$. However, we must consider the impact of the additional term $(1 - e^{\beta x})^{-\gamma}$ on this behavior, as it has the potential to alter the linear and constant response based on the values of the parameters $\beta$ and $\gamma$. We can call this parameter "Marketing Sensitivity Index" as describes how consumer response is affected by changes in advertising expenditure.

\item The parameter $C$ represents the overall effectiveness or impact multiplier of the advertising channel. It reflects the inherent effectiveness of the channel in generating consumer response, irrespective of the advertising spend, while the channel influence obtained from attribution models measures the overall effectiveness of a particular advertising channel. Higher $C$ values indicate channels with greater inherent effectiveness. A suitable designation for parameter alpha could be "Response Scaling Factor" or "Marketing Effectiveness Index". These designations highlight the importance of alpha in scaling the relationship between advertising spending and consumer response. Parameter $C$ could be referred to as "Baseline Effectiveness Constant" or "Consumer Response Baseline." These designations highlight $C$'s role in representing the inherent effectiveness of an advertising channel or campaign, regardless of advertising spend.

\item The expression $(1 - e^{\beta x})$ introduces a non-linear adjustment to the scaling behavior introduced by $Cx^{\alpha}$, incorporating the symmetry-breaking effects observed in phase transitions. Symmetry-breaking occurs when a system transitions from a symmetric state to an asymmetric one, often triggered by external factors or critical thresholds. This term is equivalent to a partition function as shown in section 2.2
As advertising spend increases $x$, this term gradually shifts the curve, reflecting changes in consumer behavior. In our equation, beta represents the rate of change in consumer response with respect to advertising spend and does not illustrate how the change evolves with increasing investment directly. Instead, beta quantifies the sensitivity or responsiveness of consumer response to changes in advertising spend. A positive beta value indicates that consumer response changes rapidly in response to variations in advertising spend, while a negative beta value suggests that consumer response will reach a saturation point from where response $y$ does not increase when increasing spend $x$. The parameter can be referred to as "Response Sensitivity Coefficient" as highlight the role of beta in quantifying the sensitivity of consumer response to changes in advertising spend. Beta can have positive and negative values. For $\beta < 0$ the term $Cx^{\alpha}(1 - e^{\beta x})$ will have a decreasingly growing behavior with a saturation level with the term $e^{\beta x}$ illustrating an exponential decay. Furthermore, it is important to note that the variable $\beta$ cannot have a value of zero, as this would result in the equation $y = 0$ for all values of $x$. For $\beta > 0$ consumer response does not show a diminishing returns effect.

\item Gamma, $\gamma$ represents the critical exponent characterizing the divergence of the correlation length of consumer behavior (scaling hypotheses near a critical point Equation 12). This parameter quantifies how the correlation length of consumer behavior changes as critical thresholds are approached. A higher $\gamma$ (more negative) value indicates a more rapid increase in the correlation length of consumer behavior as critical points are reached, suggesting that consumer behavior becomes more interconnected and influenced by external factors as critical thresholds are approached. A larger correlation length indeed suggests that the impact of the advertising campaign extends over a larger group or cluster of consumers due to a shift from individual responses to the emergence of interconnected audience clusters (critical point). Conversely, if its value is low, it suggests that customers behave more independently and disengaged from one another than collectively. When considering the correlation length in consumer response dynamics we can establish similarities with percolation theory \cite{Essam1980}. Large correlation length indicates that the influence of the advertising campaign reaches a wider range of interconnected consumers (large audience), resembling the way percolation theory explains the emergence of large connected clusters in random systems. For parameter $\gamma$, a suitable designation could be "Response Dynamics Exponent" or "Behavioral Sensitivity Index." These designations emphasize the role of $\gamma$ in quantifying the responsiveness and dynamics of consumer behavior to changes in advertising stimuli. Since gamma ($\gamma$) is the exponent of the term $(1 - e^{\beta x})$, when $\gamma = 0$, $(1 - e^{\beta x})^{\gamma} = 1$ and the consumer response curve will have the form $y = Cx^{\alpha}$ been symmetric and following a scaling law. This will lead to a concave shape growing curve with no saturation point.
\end{itemize}

It is important to note here that saturation point is not the same than tipping points. Tipping point refers to a critical spend quantity ($x$) meanwhile the critical point refers to consumer interaction amount. To find the tipping point where the curve transition from a symmetric state to an asymmetric one we have to calculate the first derivative of function $f(x)$. We can write this derivative as follows:

\begin{equation}
\frac{dy}{dx} = C\alpha x^{\alpha-1}(1 - e^{\beta x})^{-\gamma} + \gamma\beta e^{\beta x}(1 - e^{\beta x})^{-\gamma-1}
\end{equation}

Tipping point will occur when $\frac{dy}{dx} = 0$, therefore:
\begin{equation}
0 = (1 - e^{\beta x})^{\gamma}C\alpha x^{\alpha-1} + -\gamma\beta e^{\beta x}(1 - e^{\beta x})^{\gamma-1}
\end{equation}

In order to obtain a numerical solution for the equation 18, we can use iterative techniques such as Newton's method to find the solution \cite{Battiti1992}. We can prove that equation 17 is not scale invariant. The probability distribution $p(x)$ does not follow a scaling law, and the cumulative distribution function $P(x)$ does not either.

We can rewrite our proposed equation
\begin{equation}
y = Cx^{\alpha}(1 - e^{\beta x})^{-\gamma}
\end{equation}

as $f(ax) = C(ax)^{\alpha}(1 - e^{\beta ax})^{-\gamma}$. Because $\neg\{(1 - e^{\beta x})^{\gamma} \sim (1 - e^{\beta ax})^{-\gamma}\}$, then we have 
\begin{equation}
\neg\{Cx^{\alpha}x^{\alpha}(1 - e^{\beta ax})^{-\gamma} \sim x\}.
\end{equation}
By changing the value of the function's argument ($x$), the shape of the function $y$ is not maintained due to the introduction of a symmetry-breaking effect with the term $(1 - e^{\beta x})^{-\gamma}$.

\section{Experimental work}

\subsection{Material and methods}

To validate our equation, we have used three dummy datasets containing media spend data  as independent variables with one dependent variable. First, we have used a MMM model to calculate the impact of each media on the Key Performance Indicator (KPI) as a measure of consumer response. We have used the open source library Lightweight MMM developed by Google -\url{https://github.com/google/lightweight_mmm}, with the configuration detailed in Table \ref{tab:model-params}. Based on the attribution model results, we utilized the Python library SciPy for curve fitting (see Table \ref{tab:curve-fitting}) to obtain the various parameters for each of the models described in Table \ref{tab:equations}. To validate our equation, we have fitted the data using three different equations: our proposed equation, the Michaelis-Menten model equation, and the Hill equation (Table \ref{tab:equations}). The validation will be done by comparing different metrics for the curve fitting obtained with different models (Tables 4-5-6).

\begin{table}[h]
\centering
\caption{Model parameters}
\label{tab:model-params}
\begin{tabular}{lc}
\toprule
Parameters & Values \\
\midrule
model\_name & 'carryover' \\
seasonality\_degrees & 4 \\
acceptance\_p & 0.85 \\
number\_warmup & 2000 \\
samples & 2000 \\
\bottomrule
\end{tabular}
\end{table}

\begin{table}[h]
\centering
\caption{Equations used for fitting consumer response curve}
\label{tab:equations}
\begin{tabular}{lc}
\toprule
Proposed equation & $y = Cx^{\alpha}(1 - e^{\beta x})^{-\gamma}$ \\
Michaelis-Menten equation & $y = \frac{V_{max}x}{K_m + x}$ \\
Hill's equation (Hill, 1910) & $y = \frac{1}{1 + \left(\frac{K_d}{x}\right)^n}$ \\
\bottomrule
\end{tabular}
\end{table}

\begin{table}[h]
\centering
\caption{Curve fitting method\protect\footnotemark}
\label{tab:curve-fitting}
\begin{tabular}{lc}
\toprule
Parameters & Values \\
\midrule
method & 'L-BFGS-B' \\
options: & \\
\quad'disp' & True \\
\quad'maxiter' & 1E+6 \\
tol & 1E-6 \\
\bottomrule
\end{tabular}
\end{table}
\footnotetext{using python SciPy library (\url{https://docs.scipy.org/doc/scipy/reference/generated/scipy.optimize.minimize.html})}

\subsection{Results}

Before presenting the results, we can make a quick reminder of our equation parameters:

$C$ - Consumer response baseline. 

$\alpha$- Marketing Effectiveness: Channel’s ability to scale impact with spend. It can range from 0 to 1. High values (>0.5) indicate complex, multiple mechanisms. Low values (<0.2) indicate a more focused mechanism.

$\beta$ — Response Sensitivity: How quickly a channel reaches diminishing returns. It can range from 0 to 1. High values (>0.8) imply rapid phase transition and a defined saturation point.Low values (<0.2) indicate gradual changes, with no clear saturation.

$\gamma$ — Behavioral Sensitivity: Audience clustering and viral potential It can range from 0 to 1. High values (>0.8) indicate strong network effects and a unique unified audience. Low values (<0.2) indicate that there is a more fragmented audience, each of them giving independent responses.

For example, the combination of high $\alpha$ + high $\gamma$ values reflects a complex channel with strong network effects. These kinds of channels are ideal for launching viral campaigns with a broad reach. A high $\alpha$ + low $\gamma$ suggests a complex channel with a fragmented audience. This combination is suitable for targeted campaigns with diversified messaging. The combination of low $\alpha$ + high $\gamma$ reveals a focused channel with a unified audience. This combination is suitable for precision targeting and community building.

\vspace{12pt}
Table \ref{tab:results-data1} shows the results obtained for dummy data 1. We can see, media channel Facebook has the highest consumer response baseline ($C$), which suggests that it is more effective in creating consumer response. In other words, its response curve initially grows at the highest rate. Google media channel has the lowest ($C$) value, indicating a comparatively lower level of intrinsic response when compared to other channels.

\begin{table}[H]
\centering
\caption{Equation fitting for dummy data 1}
\label{tab:results-data1}
\begin{adjustbox}{width=\textwidth}
\begin{tabular}{lcccccc}
\toprule
 & Marketing & Response & Behavioral & Consumer & & Channel \\
 & Effectiveness, $\alpha$ & Sensitivity, $\beta$ & Sensitivity, $\gamma$ & Response & RoAS & influence \\
 & & & & Baseline, $C$ & & (\%) \\
\midrule
TV spend (offline) & 0.165 & -0.072 & 0.000 & 54997.3 & 0.929 & 5.62 \\
Out-of-home & & & & & & \\
advertising spend & 0.018 & 0.286 & 0.008 & 77805.3 & 0.451 & 2.01 \\
(offline) & & & & & & \\
Print adds spend & 0.048 & -1.000 & 1.000 & 77556.3 & 1.341 & 2.07 \\
(offline) & & & & & & \\
Google search spend & 0.150 & 0.004 & 0.004 & 37877.0 & 1.865 & 4.55 \\
Facebook spend & 0.045 & -0.011 & 1.000 & 90718.6 & 0.748 & 2.64 \\
\bottomrule
\end{tabular}
\end{adjustbox}
\end{table}

TV spend channel has the highest marketing effectiveness index ($\alpha$), suggesting that small changes in TV advertising channel spending could lead to significant changes in consumer response. It means that with increasing the spend we can expect an important revenue increase. The marketing effectiveness ($\alpha$) value for the channel Out-Of-Home advertising (OOH) is the lowest, suggesting a response curve that is reasonably flat and that spend increase does not convert to revenue generation. The marketing sensitivity index ($\alpha$), does not have a linear relationship with the channel influence obtained from an attribution model due to the significant impact of the customer response baseline. For instance, a high marketing effectiveness ($\alpha$) value indicates a high level of channel importance if the consumer response baseline is also high. OOH exhibits the highest positive response sensitivity values ($\beta$), suggesting that from a certain spend level (or critical point) the outcome is almost constant. The response sensitivity ($\beta$) value for channel Print ads is the lowest (negative), indicating that consumer response achieves saturation rapidly. The behavioral sensitivity index ($\gamma$), for both channels TV and Google search has a reasonably low level, implying that we are targeting several audiences who are acting independently. This suggests the presence of several different small clusters, as per the concepts of percolation theory. For Print ads and Facebook spend, behavioral sensitivity index ($\gamma$) values are high, indicating a significant increase in the correlation length of consumer behavior near critical points. This implies that consumers are more interconnected behaving as a homogeneous audience.

\vspace{12pt}
Tables \ref{tab:results-data2} and \ref{tab:results-data3} show the results obtained for dummy data 2 and dummy data 3. As we previously discussed, we can see from these Tables that the relationship between the channel's influence and the combination of the Marketing Sensitivity Index and the Consumer Response Baseline is proportional. However, these two parameters separately do not show proportionality with the channel influence obtained from the MMM model. For example, in Table \ref{tab:results-data2} we see that channel Offline 3 has the highest influence (20.22\%). Marketing effectiveness ($\alpha$) is 0.1919, lower than channel Offline 1  has a value of 0.3434. In this case, channel Offline 1 has a consumer response baseline ($C$) of 5078.6, much lower than Offline 3 channel (13883.8).

\begin{table}[h]
\centering
\caption{Equation fitting for dummy data 2}
\label{tab:results-data2}
\begin{adjustbox}{width=\textwidth}
\begin{tabular}{lcccccc}
\toprule
 & Marketing & Response & Behavioral & Consumer & & Channel \\
 & Effectiveness, $\alpha$ & Sensitivity, & Sensitivity, & Response & RoAS & influence \\
 & & $\beta$ & $\gamma$ & Baseline, $C$ & & (\%) \\
\midrule
Online 1 & 0.228 & 0.010 & 0.075 & 5078.6 & 9.65 & 10.43 \\
Offline 1 & 0.343 & -0.164 & 0.000 & 5122.7 & 8.97 & 8.08 \\
Offline 2 & 0.041 & 0.082 & 0.223 & 13836.5 & 62.69 & 1.55 \\
Offline 3 & 0.192 & -0.012 & 0.009 & 13883.8 & 2.37 & 20.22 \\
Offline 4 & 0.034 & 0.006 & 0.145 & 35052.3 & 86.49 & 4.63 \\
Offline 5 & 0.378 & 0.002 & 0.068 & 6245.7 & 29.40 & 29.93 \\
\bottomrule
\end{tabular}
\end{adjustbox}
\end{table}

\begin{table}[H]
\centering
\caption{Equation fitting for dummy data 3}
\label{tab:results-data3}
\begin{adjustbox}{width=\textwidth}
\begin{tabular}{lcccccc}
\toprule
 & Marketing & Response & Behavioral & Consumer & & Channel \\
 & Effectiveness, & Sensitivity, $\beta$ & Sensitivity, $\gamma$ & Response & RoAS & influence \\
 & $\alpha$ & & & Baseline, $C$ & & (\%) \\
\midrule
Brand Search & 0.069 & 0.889 & 0.140 & 1.63 & 0.23 & 0.38 \\
Partnerships & 0.515 & 0.197 & 0.835 & 1.78 & 0.83 & 0.97 \\
TV & 0.667 & 0.327 & 0.028 & 1.478 & 0.35 & 2.14 \\
Programmatic & 0.458 & 0.984 & 0.298 & 2.87 & 1.17 & 2.4 \\
Magazines 1 & 0.636 & -0.008 & 0.000 & 3.75 & 1.72 & 1.04 \\
Magazines 2 & 0.179 & -0.999 & 0.935 & 11.67 & 6.64 & 0.95 \\
Magazines 3 & 0.141 & 0.854 & 0.299 & 41.07 & 11.57 & 5.76 \\
Advertorials & 0.39 & 0.423 & 0.325 & 1.47 & 0.35 & 1.39 \\
Sponsorship & 0.594 & 0.061 & 0.004 & 3.62 & 1.22 & 1.99 \\
Business Events 1 & 0.686 & 0.059 & 0 & 9.52 & 3.40 & 2.62 \\
Business Events 2 & 0.671 & -0.017 & 0.009 & 17.31 & 6.11 & 4.01 \\
Business Events 3 & 0.337 & 0.002 & 0.000 & 2.49 & 0.44 & 0.51 \\
OOH & 0.492 & 0.308 & 0.216 & 1.935 & 0.37 & 1.53 \\
\bottomrule
\end{tabular}
\end{adjustbox}
\end{table}

\begin{figure}[H]
\centering
\includegraphics[height=0.92\textheight]{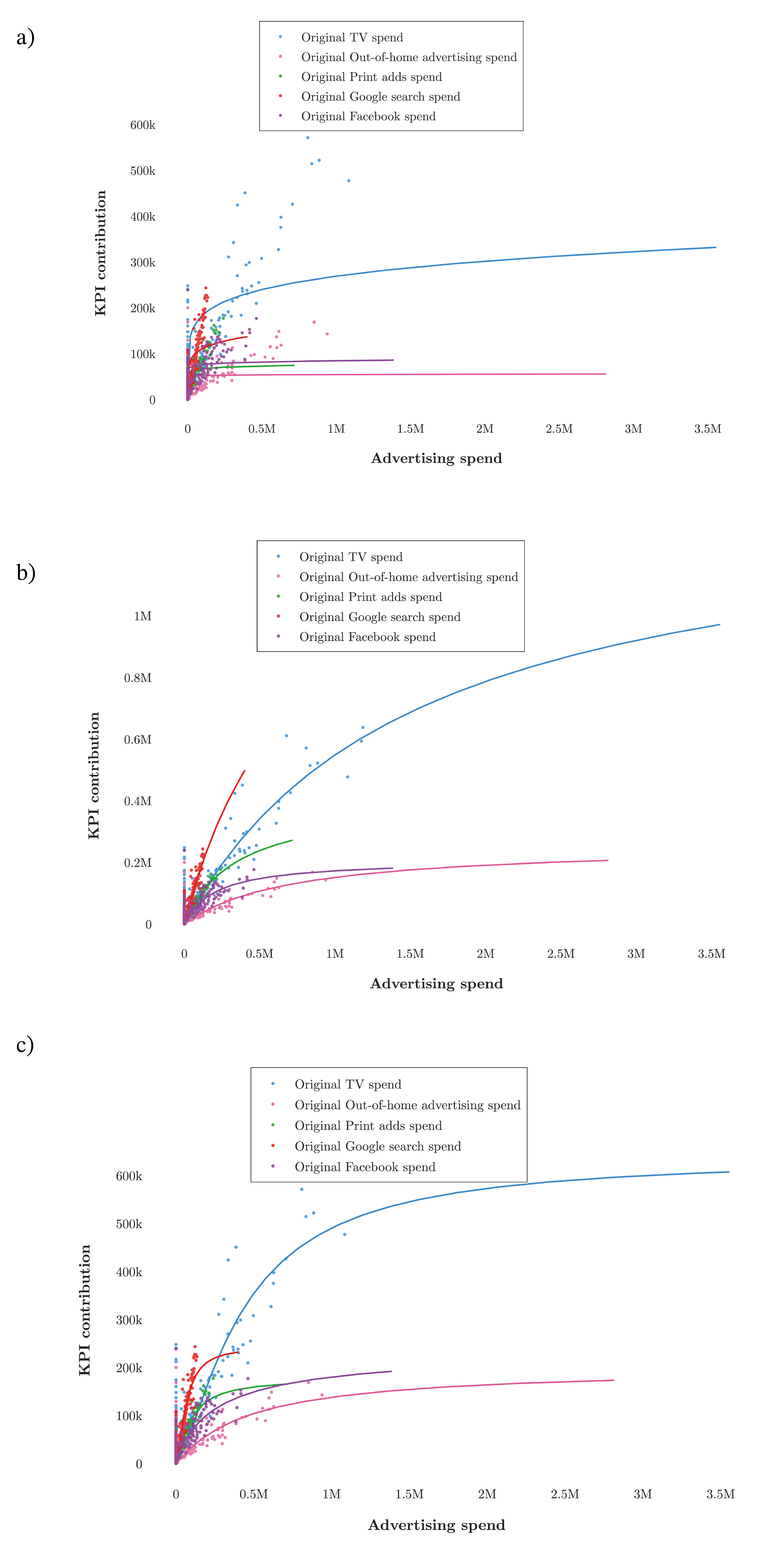}
\caption{Curves obtained with dummy data 1. a) Proposed equation, b) Michaelis-Menten equation, c) Hill's equation}
\label{fig:curves-data1}
\end{figure}

\begin{figure}[H]
\centering
\includegraphics[height=0.92\textheight]{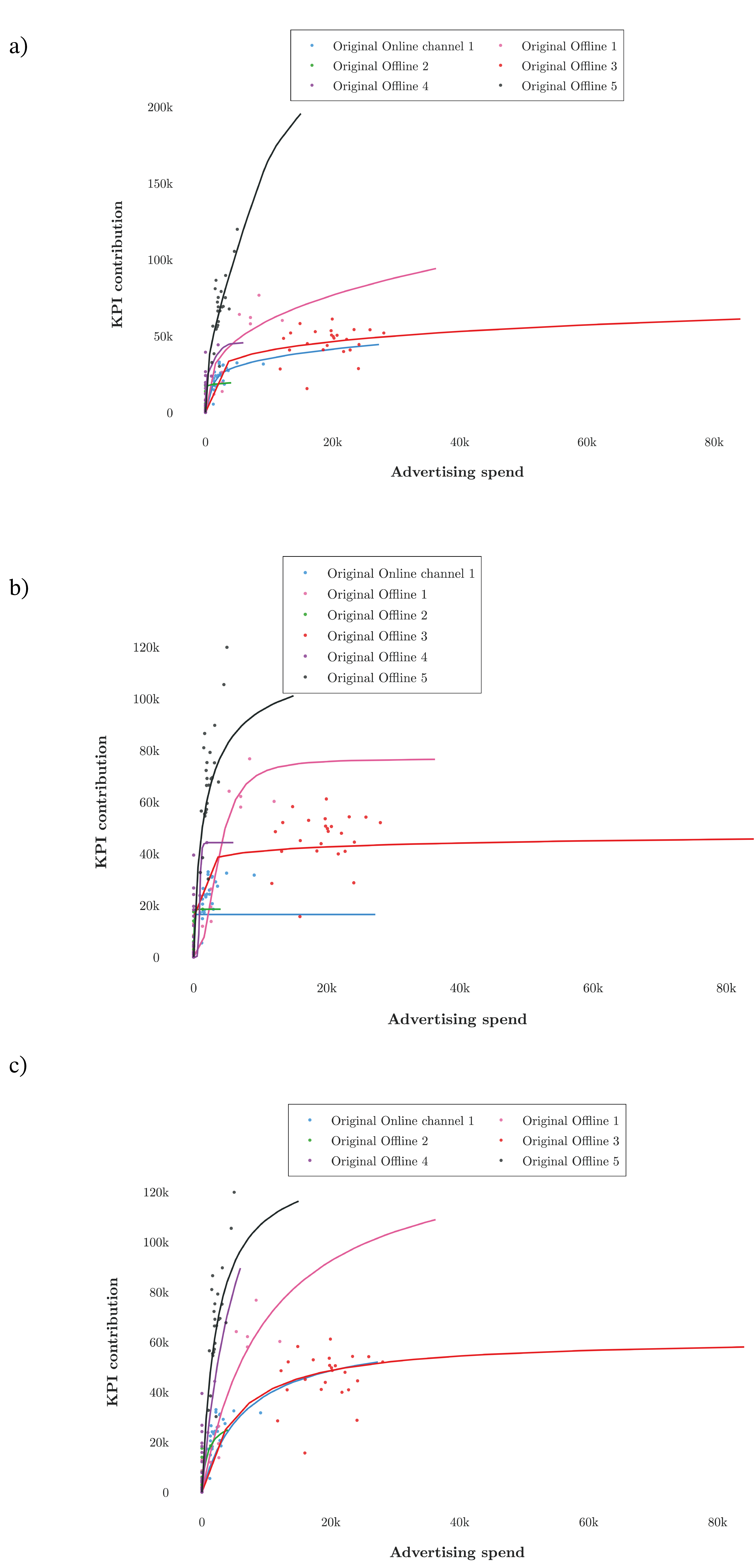}
\caption{Curves obtained with dummy data 2. a) Proposed equation, b) Michaelis-Menten equation, c) Hill's equation}
\label{fig:curves-data2}
\end{figure}

\begin{figure}[H]
\centering
\includegraphics[height=0.92\textheight]{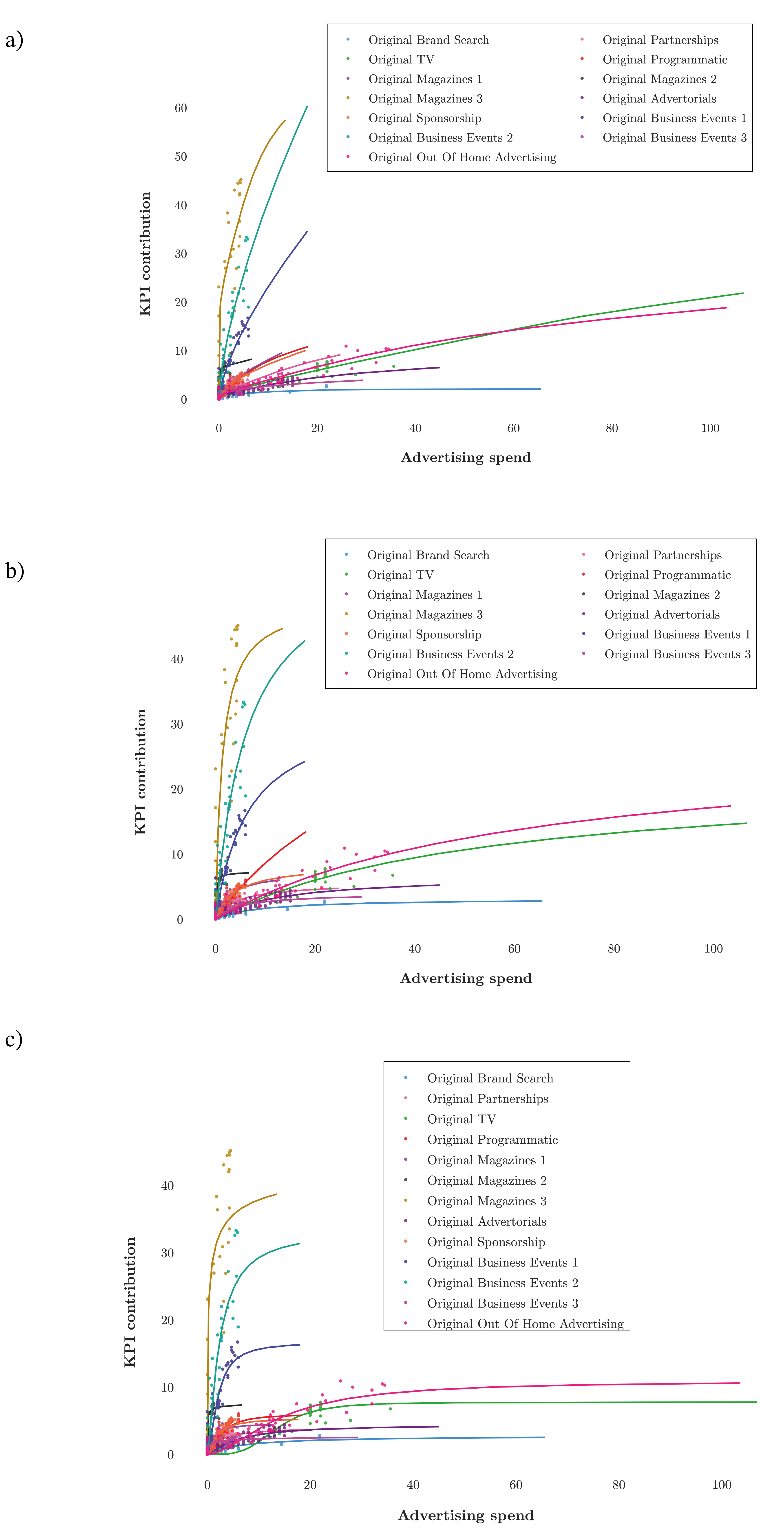}
\caption{Curves obtained with dummy data 3. a) Proposed equation, b) Michaelis-Menten equation, c) Hill's equation}
\label{fig:curves-data3}
\end{figure}

The curves obtained from applying the proposed equation to our dummy datasets are shown in Figures \ref{fig:curves-data1}, \ref{fig:curves-data2} and \ref{fig:curves-data3}. The different Tables show the value of Response Sensitivity ($\beta$) is not consistently negative. A negative beta value implies a diminishing returns effect, where the curve first increases rapidly (lower $x$) but gradually slows down until it approaches saturation.

Positive values of response sensitivity ($\beta$) indicate the absence of a saturation point, and the curve will continue to grow to a certain extent based on that value without decreasing its growth rate. Unlike the other models used, our equation does not necessarily imply that the consumer's response will saturate. If we wanted that to be the case, we would simply need to ensure that the value of beta is always negative. However, it has been deemed interesting for this value to have the possibility of not saturating, with the advantage that by not forcing saturation (as is the case with the Michaelis-Menten equation with $y = V_{max}$), the response curve will grow to a lesser extent and be more realistic. This is clearly illustrated in Figures \ref{fig:curves-data1}, \ref{fig:curves-data2}, and \ref{fig:curves-data3}. The curves obtained using the Hill and Michaelis-Menten equations grow to far higher values than those obtained with the proposed equation. This inevitably leads to a poorer fit of the curve at small spend values ($x$), as these other equations primarily focus on fitting higher values of the response ($y$) in order to find the saturation point. Our equation solves this problem by incorporating a term equal to a partition function with a parameter that has a similar effect to the Boltzmann factor (equation 6).

The Behavioral Sensitivity Index (gamma) values exhibit a significant range, spanning from 0 to 1. The impact of this parameter on the response curve is not straightforward and needs to be evaluated in relation to percolation theory. Consider a network in which consumers are represented as nodes, and the connections between them represent interactions or influences related to advertising. Percolation theory can be applied to study the propagation of advertising messages through connections and their impact on customer behavior.

In equation 16, the parameter Behavioral Sensitivity Index ($\gamma$) quantifies the behavior of our system in proximity to this threshold. Below this threshold, there are clusters of connected nodes that are not sufficiently large to cover the full network. When the threshold is exceeded, a large interconnected cluster forms, covering a substantial part of the network. Regarding advertising, this threshold signifies the minimum amount of exposure or acceptance required for a specific product or concept to achieve broad popularity within a population. When consumers engage with advertising messages, they have the potential to create groups of influence that are based on common qualities, preferences, or social relationships. These clusters denote cohorts of consumers who are inclined to adopt comparable attitudes or behaviors in reaction to advertising. Cluster formation occurs when particular advertising efforts powerfully resonate with specific audiences.

In our experiments, the value of behavioral sensitivity index ($\gamma$) illustrates if our audience is fragmented or if it's acting as a large group. Channels with larger behavioral sensitivity index ($\gamma$) values can include mainstream television channels, popular social media platforms, widely-read newspapers or magazines, and influential websites, and has the potential of easily spread brand messages.

Table \ref{tab:metrics-data1} shows the results of using an Ordinary Least Squares (OLS) regression analysis including an intercept term (standard regression), whereas the subsequent Table showcases the findings obtained when the intercept term is excluded (Restricted Total OLS regression, RTO). Channels are in order: TV, OOH, Print ads., Google search and Facebook.

\begin{table}[h]
\centering
\caption{Metrics for curve fitting in dummy data 1.}
\label{tab:metrics-data1}
\begin{adjustbox}{width=\textwidth}
\begin{tabular}{lccccccccc}
\toprule
& \multicolumn{3}{c}{Hill's equation} & \multicolumn{3}{c}{Michaelis-Menten} & \multicolumn{3}{c}{Proposed equation} \\
\cmidrule(lr){2-4} \cmidrule(lr){5-7} \cmidrule(lr){8-10}
& $r^2$ & p-val & F-pval & $r^2$ & p-val & F-pval & $r^2$ & p-val & F-pval \\
\midrule
\multicolumn{10}{l}{With intercept:} \\
Ch1 & 0.000 & 0.955 & 0.955 & 0.001 & 0.680 & 0.680 & 0.001 & 0.728 & 0.728 \\
Ch2 & 0.087 & 0.000 & 0.000 & 0.102 & 0.000 & 0.000 & 0.002 & 0.546 & 0.546 \\
Ch3 & 0.042 & 0.003 & 0.003 & 0.043 & 0.003 & 0.003 & 0.041 & 0.003 & 0.003 \\
Ch4 & 0.346 & 0.000 & 0.000 & 0.356 & 0.000 & 0.000 & 0.258 & 0.000 & 0.000 \\
Ch5 & 0.004 & 0.363 & 0.363 & 0.004 & 0.373 & 0.373 & 0.009 & 0.174 & 0.174 \\
Avg & 0.096 & 0.264 & 0.264 & 0.101 & 0.211 & 0.211 & 0.062 & 0.290 & 0.290 \\
\midrule
\multicolumn{10}{l}{RTO model:} \\
Ch1 & 0.329 & 0.000 & 0.000 & 0.311 & 0.000 & 0.000 & 0.382 & 0.000 & 0.000 \\
Ch2 & 0.336 & 0.000 & 0.000 & 0.315 & 0.000 & 0.000 & 0.501 & 0.000 & 0.000 \\
Ch3 & 0.359 & 0.000 & 0.000 & 0.337 & 0.000 & 0.000 & 0.495 & 0.000 & 0.000 \\
Ch4 & 0.785 & 0.000 & 0.000 & 0.787 & 0.000 & 0.000 & 0.741 & 0.000 & 0.000 \\
Ch5 & 0.471 & 0.000 & 0.000 & 0.470 & 0.000 & 0.000 & 0.554 & 0.000 & 0.000 \\
Avg & 0.456 & 0.000 & 0.000 & 0.444 & 0.000 & 0.000 & 0.535 & 0.000 & 0.000 \\
\bottomrule
\end{tabular}
\end{adjustbox}
\end{table}

Table \ref{tab:metrics-data2} shows the results of using an Ordinary Least Squares (OLS) regression analysis including an intercept term (standard regression), whereas the subsequent Table showcases the findings obtained when the intercept term is excluded (Restricted Total OLS regression, RTO). Channels are in order: Online 1, Offline 1, Offline 2, Offline 3, Offline 4 and Offline 5.

\begin{table}[h]
\centering
\caption{Metrics for curve fitting in dummy data 2.}
\label{tab:metrics-data2}
\begin{adjustbox}{width=\textwidth}
\begin{tabular}{lccccccccc}
\toprule
& \multicolumn{3}{c}{Hill's equation} & \multicolumn{3}{c}{Michaelis-Menten} & \multicolumn{3}{c}{Proposed equation} \\
\cmidrule(lr){2-4} \cmidrule(lr){5-7} \cmidrule(lr){8-10}
& $r^2$ & p-val & F-pval & $r^2$ & p-val & F-pval & $r^2$ & p-val & F-pval \\
\midrule
\multicolumn{10}{l}{With intercept:} \\
Ch1 & -inf & 0.000 & 1.000 & 0.033 & 0.396 & 0.396 & 0.096 & 0.140 & 0.140 \\
Ch2 & 0.064 & 0.234 & 0.234 & 0.140 & 0.071 & 0.071 & 0.101 & 0.131 & 0.131 \\
Ch3 & 0.016 & 0.558 & 0.558 & 0.338 & 0.003 & 0.003 & 0.002 & 0.846 & 0.846 \\
Ch4 & 0.551 & 0.000 & 0.000 & 0.548 & 0.000 & 0.000 & 0.555 & 0.000 & 0.000 \\
Ch5 & 0.138 & 0.074 & 0.074 & 0.057 & 0.260 & 0.260 & 0.016 & 0.559 & 0.559 \\
Avg & 0.334 & 0.003 & 0.003 & 0.318 & 0.004 & 0.004 & 0.319 & 0.004 & 0.004 \\
\midrule
\multicolumn{10}{l}{RTO model:} \\
Ch1 & 0.926 & 0.000 & 0.000 & 0.851 & 0.000 & 0.000 & 0.893 & 0.000 & 0.000 \\
Ch2 & 0.388 & 0.001 & 0.001 & 0.447 & 0.000 & 0.000 & 0.419 & 0.000 & 0.000 \\
Ch3 & 0.306 & 0.004 & 0.004 & 0.089 & 0.147 & 0.147 & 0.278 & 0.007 & 0.007 \\
Ch4 & 0.975 & 0.000 & 0.000 & 0.976 & 0.000 & 0.000 & 0.972 & 0.000 & 0.000 \\
Ch5 & 0.089 & 0.147 & 0.147 & 0.216 & 0.019 & 0.019 & 0.355 & 0.002 & 0.002 \\
Avg & 0.945 & 0.000 & 0.000 & 0.939 & 0.000 & 0.000 & 0.903 & 0.000 & 0.000 \\
\bottomrule
\end{tabular}
\end{adjustbox}
\end{table}

Table \ref{tab:metrics-data3} shows the results of using an Ordinary Least Squares (OLS) regression analysis including an intercept term (standard regression), whereas the subsequent Table showcases the findings obtained when the intercept term is excluded (Restricted Total OLS regression, RTO). Channels are in order: Brand Search, Partnerships, TV, Programmatic, Magazines 1, Magazines 2, Magazines 3, Advertorials, Sponsorship, Business Events 1, Business Events 2, Business Events 3 and Out Of Home Advertising.

\begin{table}[H]
\centering
\caption{Metrics for curve fitting in dummy data 3.}
\label{tab:metrics-data3}
\begin{adjustbox}{width=\textwidth}
\begin{tabular}{lccccccccc}
\toprule
& \multicolumn{3}{c}{Hill's equation} & \multicolumn{3}{c}{Michaelis-Menten} & \multicolumn{3}{c}{Proposed equation} \\
\cmidrule(lr){2-4} \cmidrule(lr){5-7} \cmidrule(lr){8-10}
& $r^2$ & p-val & F-pval & $r^2$ & p-val & F-pval & $r^2$ & p-val & F-pval \\
\midrule
\multicolumn{10}{l}{With intercept:} \\
Ch1 & 0.000 & 0.842 & 0.842 & 0.000 & 0.922 & 0.922 & 0.000 & 0.874 & 0.874 \\
Ch2 & 0.209 & 0.000 & 0.000 & 0.182 & 0.000 & 0.000 & 0.147 & 0.000 & 0.000 \\
Ch3 & 0.006 & 0.446 & 0.446 & 0.000 & 0.943 & 0.943 & 0.000 & 0.958 & 0.958 \\
Ch4 & 0.037 & 0.049 & 0.049 & 0.042 & 0.037 & 0.037 & 0.047 & 0.026 & 0.026 \\
Ch5 & 0.093 & 0.002 & 0.002 & 0.075 & 0.005 & 0.005 & 0.063 & 0.010 & 0.010 \\
Ch6 & 0.002 & 0.669 & 0.669 & 0.003 & 0.590 & 0.590 & 0.003 & 0.575 & 0.575 \\
Ch7 & 0.105 & 0.001 & 0.001 & 0.155 & 0.000 & 0.000 & 0.153 & 0.000 & 0.000 \\
Ch8 & 0.048 & 0.024 & 0.024 & 0.046 & 0.028 & 0.028 & 0.042 & 0.037 & 0.037 \\
Ch9 & 0.054 & 0.017 & 0.017 & 0.087 & 0.002 & 0.002 & 0.110 & 0.001 & 0.001 \\
Ch10 & 0.092 & 0.002 & 0.002 & 0.097 & 0.001 & 0.001 & 0.100 & 0.001 & 0.001 \\
Ch11 & 0.194 & 0.000 & 0.000 & 0.212 & 0.000 & 0.000 & 0.210 & 0.000 & 0.000 \\
Ch12 & 0.205 & 0.000 & 0.000 & 0.372 & 0.000 & 0.000 & 0.365 & 0.000 & 0.000 \\
Ch13 & 0.003 & 0.603 & 0.603 & 0.005 & 0.484 & 0.484 & 0.005 & 0.465 & 0.465 \\
Avg & 0.081 & 0.204 & 0.204 & 0.098 & 0.232 & 0.232 & 0.096 & 0.227 & 0.227 \\
\midrule
\multicolumn{10}{l}{RTO model:} \\
Ch1 & 0.321 & 0.000 & 0.000 & 0.312 & 0.000 & 0.000 & 0.332 & 0.000 & 0.000 \\
Ch2 & 0.349 & 0.000 & 0.000 & 0.329 & 0.000 & 0.000 & 0.332 & 0.000 & 0.000 \\
Ch3 & 0.308 & 0.000 & 0.000 & 0.386 & 0.000 & 0.000 & 0.392 & 0.000 & 0.000 \\
Ch4 & 0.792 & 0.000 & 0.000 & 0.788 & 0.000 & 0.000 & 0.788 & 0.000 & 0.000 \\
Ch5 & 0.324 & 0.000 & 0.000 & 0.330 & 0.000 & 0.000 & 0.338 & 0.000 & 0.000 \\
Ch6 & 0.175 & 0.000 & 0.000 & 0.167 & 0.000 & 0.000 & 0.174 & 0.000 & 0.000 \\
Ch7 & 0.251 & 0.000 & 0.000 & 0.301 & 0.000 & 0.000 & 0.274 & 0.000 & 0.000 \\
Ch8 & 0.490 & 0.000 & 0.000 & 0.484 & 0.000 & 0.000 & 0.508 & 0.000 & 0.000 \\
Ch9 & 0.527 & 0.000 & 0.000 & 0.500 & 0.000 & 0.000 & 0.502 & 0.000 & 0.000 \\
Ch10 & 0.140 & 0.000 & 0.000 & 0.141 & 0.000 & 0.000 & 0.144 & 0.000 & 0.000 \\
Ch11 & 0.087 & 0.002 & 0.002 & 0.080 & 0.003 & 0.003 & 0.086 & 0.002 & 0.002 \\
Ch12 & 0.362 & 0.000 & 0.000 & 0.484 & 0.000 & 0.000 & 0.435 & 0.000 & 0.000 \\
Ch13 & 0.216 & 0.000 & 0.000 & 0.214 & 0.000 & 0.000 & 0.222 & 0.000 & 0.000 \\
Avg & 0.334 & 0.000 & 0.000 & 0.347 & 0.000 & 0.000 & 0.348 & 0.000 & 0.000 \\
\bottomrule
\end{tabular}
\end{adjustbox}
\end{table}

Tables \ref{tab:metrics-data1}, \ref{tab:metrics-data2} and \ref{tab:metrics-data3} represent the outcome of an Ordinary Least Squares (OLS) regression with an intercept (standard regression) and without an intercept, known as the Restricted Total OLS (RTO) regression. It is important to note that when $x$ is equal to zero, $y$ should also be equal to zero. We assume that there is no threshold in the variable $y$ as discussed in section 1. This implies that when the spend is zero, the response is also zero. But this information alone is not enough to completely explain why regression through the origin (RTO) is a more suitable fit. The primary distinction between the two models lies in the fact that the regression via the origin model has a higher coefficient of determination ($r^2$). A higher value of $r^2$ does not necessarily indicate a superior fit, because it could be possible the model is overfit.

Based on the $r^2$ values in our RTO models, it appears unlikely that the model is overfit in any of the datasets (only in dataset 2 we reach values of $r^2 > 0.9$ for Ch4 and Ch6). Furthermore, it is evident that while the predictor coefficient in the RTO model consistently shows statistical significance ($p < 0.01$), it is not necessarily statistically significant in all the regular OLS models.

\begin{table}[H]
\centering
\caption{Statistics for a standard OLS model with dummy data 3}
\label{tab:stats-ols}
\begin{adjustbox}{width=\textwidth}
\begin{tabular}{lccccccccc}
\toprule
& $r^2$ & Intercept & Coef & Int. std & Coef. std & Int. p- & Coef. & F-pval & Res. Std \\
& & & & error & error & val & p-val & & error \\
\midrule
\midrule
Ch1 & 0.000 & 4.12 & -0.02 & 0.11 & 0.125 & 0.000 & 0.874 & 0.874 & 0.947 \\
Ch2 & 0.147 & 12.51 & -1.54 & 0.53 & 0.401 & 0.000 & 0.000 & 0.000 & 4.147 \\
Ch3 & 0.000 & 41.12 & -0.03 & 2.63 & 0.645 & 0.000 & 0.958 & 0.958 & 19.408 \\
Ch4 & 0.047 & 13.91 & 1.320 & 1.93 & 0.558 & 0.000 & 0.026 & 0.026 & 7.210 \\
Ch5 & 0.063 & 14.54 & -1.0 & 0.72 & 0.469 & 0.000 & 0.010 & 0.010 & 5.447 \\
Ch6 & 0.003 & 49.77 & -0.14 & 0.80 & 0.129 & 0.000 & 0.575 & 0.575 & 6.655 \\
Ch7 & 0.153 & 1,258.0 & 8.32 & 33.28 & 1.098 & 0.000 & 0.000 & 0.000 & 273.557 \\
Ch8 & 0.042 & 11.46 & -0.69 & 0.67 & 0.334 & 0.000 & 0.037 & 0.037 & 4.076 \\
Ch9 & 0.110 & 21.49 & -1.49 & 1.58 & 0.468 & 0.000 & 0.001 & 0.001 & 6.857 \\
Ch10 & 0.100 & 175.64 & -4.48 & 7.71 & 1.487 & 0.000 & 0.001 & 0.001 & 69.030 \\
Ch11 & 0.210 & 641.96 & -13.5 & 29.14 & 1.723 & 0.000 & 0.000 & 0.000 & 227.343 \\
Ch12 & 0.365 & 5.16 & 0.96 & 0.16 & 0.093 & 0.000 & 0.000 & 0.000 & 1.255 \\
Ch13 & 0.005 & 64.64 & -0.72 & 3.27 & 0.980 & 0.000 & 0.465 & 0.465 & 29.329 \\
\bottomrule
\end{tabular}
\end{adjustbox}
\end{table}

Last, an additional test can be performed by comparing the standard error of the coefficients for both models (Tables \ref{tab:stats-ols} and \ref{tab:stats-rto}). As we can see, the coefficient values for the RTO model are considerably higher (and positive) than those obtained using the normal OLS model. As a result, the standard error for the RTO model is also smaller, since the standard error values for both models are relatively similar. For example, in Ch1 the coefficient result for the typical ordinary least squares (OLS) regression is -0.02, with a corresponding standard error of 0.125. However, the RTO model yields a larger value for Ch1's coefficient, which is 3.050 with a standard error of 0.379 (Table \ref{tab:stats-rto}).

\begin{table}[H]
\centering
\caption{Statistics for RTO model with dummy data 3}
\label{tab:stats-rto}
\begin{adjustbox}{width=\textwidth}
\begin{tabular}{lccccccc}
\toprule
 & $r^2$ & Coef & Coef. std error & p-value & F-pvalue & Res. Std error \\
\midrule
Ch1 & 0.332 & 3.05 & 0.379 & 0.000 & 0.000 & 2.185 \\
Ch2 & 0.332 & 4.04 & 0.457 & 0.000 & 0.000 & 7.487 \\
Ch3 & 0.392 & 7.40 & 0.501 & 0.000 & 0.000 & 28.266 \\
Ch4 & 0.788 & 5.30 & 0.258 & 0.000 & 0.000 & 8.685 \\
Ch5 & 0.338 & 4.53 & 0.503 & 0.000 & 0.000 & 9.083 \\
Ch6 & 0.174 & 7.59 & 0.261 & 0.000 & 0.000 & 20.370 \\
Ch7 & 0.274 & 45.09 & 2.570 & 0.000 & 0.000 & 580.797 \\
Ch8 & 0.508 & 3.71 & 0.298 & 0.000 & 0.000 & 6.767 \\
Ch9 & 0.502 & 4.60 & 0.282 & 0.000 & 0.000 & 11.969 \\
Ch10 & 0.144 & 11.03 & 1.329 & 0.000 & 0.000 & 105.336 \\
Ch11 & 0.086 & 18.33 & 1.960 & 0.002 & 0.002 & 358.377 \\
Ch12 & 0.435 & 3.36 & 0.156 & 0.000 & 0.000 & 2.671 \\
Ch13 & 0.222 & 9.39 & 0.705 & 0.000 & 0.000 & 41.673 \\
\bottomrule
\end{tabular}
\end{adjustbox}
\end{table}

After evaluating the appropriateness of the RTO model, we can conclude that the curves fitted using our suggested equation (equation 17) have a superior match in all datasets. The proposed equation yields an average $r^2$ value of 0.535 for dummy data 1, while the Hill and Michaelis-Menten models yield average values of 0.456 and 0.444, respectively. The average $r^2$ values for dummy data 2 are as follows: 0.903 for the suggested equation, and 0.945 and 0.939 for the Hill and Michaelis-Menten models, respectively. Finally, for dummy data 3, the average values of $r^2$ are 0.348 for the suggested equation and 0.334 and 0.347 for the Hill and Michaelis-Menten model, respectively. 

The discrepancy in the accuracy of the three provided equations might be traced to certain characteristics of the input data. Figures \ref{fig:curves-data1}, \ref{fig:curves-data2}, and \ref{fig:curves-data3}, especially the first two, show an important amount of data points down the y-axis at $x = 0$. Both the Michaelis-Menten and Hill equations offer an accurate method of adjusting the data while ignoring any $y$ points corresponding to $x = 0$. On the other hand, our proposed equation effectively incorporates them, showing a distinct curve shape. The curve generated by the proposed equation first shows nearly perpendicular growth until it reaches a certain threshold, at which point it takes on a convex shape. However, the Hill and Michaelis-Menten equations do not allow for a curve with this shape able to capture all these data points in the y-axis when $x = 0$. If we had used a standard OLS regression model for comparison (where $y \neq 0$ when $x = 0$) the Michaelis-Menten equation should yield better results for all datasets.

\section{Conclusions and future work}

In this study, we explored the dynamics of consumer response to advertising stimuli through the lens of various mathematical frameworks borrowed from physics and social psychology. By integrating these frameworks into our analysis, we aimed to better understand the intricate relationship between advertising inputs and consumer behavior. Our investigation revealed several key insights:

By leveraging principles from physics along with concepts from social psychology, we were able to construct a comprehensive theoretical framework for analyzing consumer responses to advertising. This framework offers a valuable perspective on understanding the underlying dynamics of audience behavior. We have introduced a novel equation to model the relationship between advertising spending and consumer response and validated it against common equations such as the Michaelis-Menten and Hill equations. Through the validation process, we proved the efficacy of our proposed equation in capturing the details of consumer response dynamics. Our analysis highlighted the importance of new parameters, such as marketing effectiveness, response sensitivity, and behavioral sensitivity in shaping the dynamics of consumer response. These parameters provide very interesting insights apart from the classical channel influence and saturation point.

The findings from this research have practical implications for advertisers and marketers. By understanding the relation between advertising spending and consumer behavior, marketers can optimize their advertising strategies to maximize the impact on key performance indicators (KPIs) and return on advertising spend (RoAS). Moreover, our insights into the behavior of different advertising channels can inform allocation decisions and resource quota for advertising campaigns.

Despite our contributions, it is important to acknowledge its limitations. The dummy datasets used for validation may not fully capture the complexity of real-world advertising scenarios. Future research could involve the analysis of actual advertising data to further validate and refine our approach.

\bibliographystyle{plainnat}

\end{document}